\documentclass[aps, 11pt, notitlepage, a4paper, fleqn, showpacs, floatfix, superscriptaddress, longbibliography]{revtex4-1}

\usepackage{silence}
\usepackage{latexsym}
\usepackage[utf8]{inputenc}
\usepackage[dvips]{color}
\usepackage[toc]{appendix}
\usepackage{titlesec,titletoc,ntheorem-hyper}       
\usepackage{graphicx,microtype,booktabs,cleveref,amsmath,fancyhdr,wrapfig,array,geometry} 

\newcommand{\be}{\begin{equation}}
\newcommand{\ee}{\end{equation}}
\newcommand{\bea}{\begin{eqnarray}}
\newcommand{\eea}{\end{eqnarray}}

\begin{document}
\title{Truncated Calogero-Sutherland Models on a Circle}
\author{Tarun R. Tummuru}
\affiliation{Department of Physics, Shiv Nadar University, Gautam Buddha Nagar, U.P. 201314, India}
\author{Sudhir R. Jain}
\affiliation{Nuclear Physics Division, Bhabha Atomic Research Centre, Mumbai 400085, India}
\author{Avinash Khare}
\affiliation{Department of Physics, University of Pune, Pune 411007, India}
\date{\today}

\vspace{0.5in}
\begin{abstract}
\textbf{Abstract.}
We investigate a quantum many-body system with particles moving on a circle and subject to two-body and three-body potentials. In this new class of models, that extrapolates from the celebrated Calogero-Sutherland model and a system with interactions among nearest and next-to-nearest neighbors, the interactions can be tuned as a function of range. We determine the exact ground state energy and wavefunction and obtain a part of the excitation spectrum. 
\end{abstract}

\maketitle 

\section{Introduction}

Exactly solvable quantum many-body problems are among the well studied fundamental systems in physics. They have been shown to elucidate the underlying general theories and help us study more complex cases. A marked example, for instance, is the application of exactly solved spin chains in the study of supersymmetric Yang-Mills theories \cite{mz}. Among these many-body systems the Calogero-Sutherland model (CSM), describing a one-dimensional many-body system with inverse square interactions, and its variants are the most prominent \cite{ha, sut, kuk}.  
	
Calogero-Sutherland systems have found applications in disparate branches of physics, ranging from quantum fractional Hall effect \cite{hal} to conformal field theories \cite{clz, car, ber, mas, cck}. They have played a crucial role in developing an understanding of subjects like generalized exclusion statistics \cite{wu,mus} and integrability of systems with long range interactions \cite{sog,sut0}. Quite interestingly, they have also been studied in conjunction with black hole physics \cite{git}. The relationship between random matrix theory (RMT) and CSM has been well explored \cite{meh,srjain}. Dyson's Brownian motion model \cite{dys} links RMT and exactly solvable models. Recent studies have expanded the list to quantum heat engines \cite{jbc, bjc} and quantum decay of unstable one-dimensional Bose gases \cite{cam}.


While Calogero \cite{cal} considered a many-body bosonic system on full line with or without harmonic confinement, Sutherland \cite{sut1,sut} in 1971 considered the corresponding problem with periodic boundary conditions and particles being constrained to move on a circle. One of the characteristic features of the CSM is that all the particles interact with each other. Nearly three decades later, two of us (Jain and Khare) \cite{jk} discovered that this system remains integrable when the interaction is limited to nearest-neighbors provided one includes a three-body attractive term.

Recently, Pittman et al. \cite{pboc} have extended the results of \cite{jk} by considering a family of one dimensional systems, on full line with harmonic confinement, in which the tunable inverse-square interactions extend over a finite number of neighbors. These systems involve pair-wise two-body as well as three-body interactions. It is then natural to inquire if one can extend this discussion to the periodic case. The purpose of this short article is to answer this question. In particular, we consider an $N$-body problem on a circle, with periodic boundary conditions, in which a particle only interact with $r$ of its neighbors via two-body and three-body terms. We obtain the exact zero-point energy and the ground state wavefunction of this system. As expected, the CSM and the JK models on circle are recovered as specific limits. We then proceed to characterize a part of the excitation spectrum.

\section{Hamiltonian and the ground state}

\begin{figure}
    \centering
    \includegraphics[height=3.5cm]{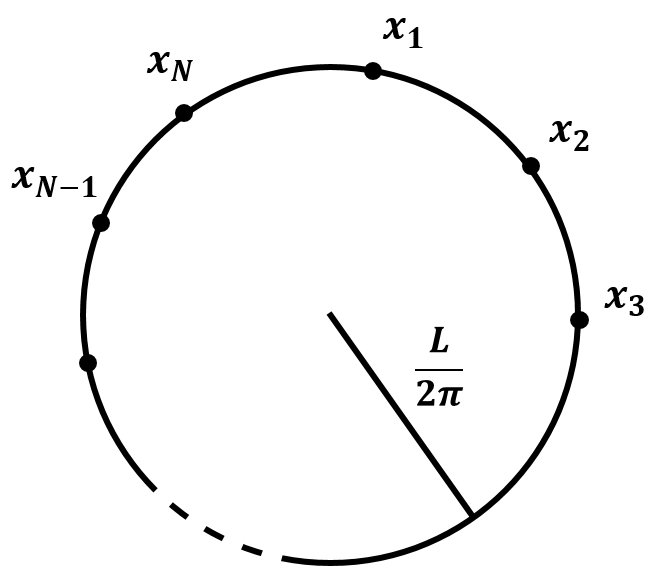}
    \caption{$N$ particles constrained to move on a circle of circumference $L$. $x_i$ ($i=1 \ldots N$) denote the coordinates measured along the circle.}
    \label{fig1}
\end{figure}

To set the notation, let us first recall that the Hamiltonian for the $N$-body problem on a circle when the range of interaction is limited to nearest neighbors is \cite{jk, ajk} 
\be\label{1} 
H = -\frac{1}{2} \sum_{i=1}^{N} \frac{\partial^2}{\partial x_i^2} 
+ g \frac{\pi^2}{L^2} \sum_{\substack{i =1}}^{N} \frac{1}{\sin ^2\left[\frac{\pi}{L} (x_{i} - x_{i+1})\right]} 
- G \frac{\pi^2}{L^2} \sum_{i=1}^{N} \cot \left[\frac{\pi}{L} 
(x_{i-1} - x_{i})\right] \cot \left[\frac{\pi}{L} (x_{i} - x_{i+1})\right],
\ee
where we have set $\hbar = m = 1$. $x_i$, $i=1 \ldots N$, denote the coordinates of the particles measured along the circle (see Fig. \ref{fig1}). and periodic boundary conditions imply $x_{N+i} = x_i$. These systems have been shown to have the ground state wave function 
\be\label{2}
\psi_0(x_1, x_2, \ldots , x_N) = \prod _{i=1}^{N} sin^{\beta} 
\left[\frac{\pi}{L} (x_i - x_{i+1})\right], 
\ee
provided $g> -\frac{1}{4}$ and $G\geq 0$ and subject to the relations
\be\label{3}
g = \beta (\beta - 1)\,,~~G = \beta^2.
\ee

We now discuss a logical extension of this problem and include interactions up to the $r^{th}$ nearest neighbor, instead of just the first nearest-neighbor. Inspired by the form of wavefunction for the CSM and JK model we start with the following ansatz for the ground state eigenfunction
\be\label{4} 
    \psi _0(x_1, x_2, \ldots , x_N) = {\mathcal N_0} \prod _{\substack{i<j \\ |i-j| \leq r}}^{N}\sin^{\beta} \left[\frac{\pi}{L} (x_i - x_{j})\right],
\ee
where ${\mathcal N_0}$ is an appropriate normalization constant. Further, we demand that this wavefunction satisfies the periodicity condition $x_{i+N} = x_{i}$. A simple way to construct it is by pairing every point on the circle with its succeeding $r$ points, proceeding either in clockwise or anti-clockwise direction (fixed in order to avoid double counting). It is worth noting that in the appropriate limits this ground state wave function interpolates between the CSM and the JK ground states. After a lengthy but straightforward algebra, the ground state \eqref{4} can be shown to be an eigenstate of the Hamiltonian
\be\label{5}
    H =  -\frac{1}{2} \sum_{i=1}^{N} \frac{\partial^2}{\partial x_i^2} + g \frac{\pi^2}{L^2} \sum_{\substack{i<j, \, \\ |i-j| \leq r}}^ {N} \frac{1}{\sin^2\left[\frac{\pi}{L} (x_{i} - x_{j})\right]} - G \frac{\pi^2}{L^2} \sum_{\substack{|i-j| \leq r \\ |j-k| \leq r \\ |k-i| > r}}^{N} \cot \left[\frac{\pi}{L} (x_{i} - x_{j})\right] \cot \left[\frac{\pi}{L} (x_{j} - x_{k})\right]. 
\ee
where, again $g> -\frac{1}{4}$ and $G\geq 0$ are related to $ \beta $ by Eq. \eqref{3}.

With $N$ and $r$ specified, we define the {\it neighborhood} of a particle as the $r$ particles that are situated on its left and right flanks. Using the notation $c = \frac{N}{2}$ or $c = \frac{N-1}{2}$, depending on whether $N$ is even or odd, observe that when $r \geq c$ any particle is in the neighborhood of every other particle and hence the system essentially goes over to the Sutherland model. 

Returning to the Hamiltonian, apart from the momenta and inverse square potentials that are inherent of the system under consideration, the constraints on the summation in the third term of Eq. (\ref{5}) are quite intriguing. They essentially convey that three particles are in an attractive potential only if all the three are not in the neighborhood of each other. In other words, a \textit{triple} chosen such that each particle falls in the neighborhood of the other two does not contribute a position-dependent potential term to the Hamiltonian. Instead, however, such a triple increases the eigenenergy by one unit. By some scrutiny we deduce that, while $r<c$, the number of these three-body terms in $H$ is given by
\be\label{6}
    \frac{N}{2} (r-k)(r+k+1)\,,
\ee
where $k$ given governed by the relations
\be\label{7}
\begin{split}
	k &= (3 r +1) - N, \qquad (2 r + 2) \leq N < (3 r + 1), \\
    &= 0, \qquad (3 r+1) \leq N.
\end{split}
\ee
The value of $k$ is irrelevant when $r \geq c$ as there are no three-body terms. \\
Further study reveals that ground state energy, noted by the following form, depends on $N$ and $r$.
\be
\begin{split}
    E_0 & = N \left[\frac{r(r+1)}{2}+\frac{k(k+1)}{6}\right]\frac{\beta ^2 \pi ^2}{L^2} \qquad r < c,\\ 
     & =  \frac{N (N^2 - 1)}{6} \frac{\beta ^2 \pi ^2}{L^2} \qquad r \geq c. 
\end{split}
\ee
As expected, in case $r =1$ the ground state energy goes over to the expression as obtained by JK \cite{jk} and to CSM \cite{sut} when $r \geq c$. Table 1 shows the energies for certain values of $N$ and $r$. 

\begin{table}[h]
    \centering
    \setlength{\tabcolsep}{15.65pt}
 	\begin{tabular}{c|c|cc}
 	\toprule 
 		$N$ & $r$ & $E_0$ \\	\midrule 
 		$6$ & $2$ & $20 \beta ^2$ \\
 		$7$ & $2$ & $21 \beta ^2$ \\
 		$8$ & $2$ & $24 \beta ^2$ \\
 		$8$ & $3$ & $56 \beta ^2$ \\
 		$9$ & $2$ & $27 \beta ^2$ \\
 		$9$ & $3$ & $30 \beta ^2$ \\	
 	\bottomrule 
 	\end{tabular} 
    \caption{With two neighbors ($r=2$), the minimum number of particles, $N$, has to be six. Here we give some typical values of ground state energies for some  systems labeled by $(N, r)$ when $r < c$.}
\end{table}

\section{Excited State Spectrum}
The eigenfunctions of the CSM Hamiltonian are expressible in terms of Jack polynomials \cite{RPS, IGM, VK}. Following this observation, Lapointe and Vinet \cite{LV} presented an elegant approach of obtaining the excited states by acting a string of creation operators on the ground state wavefunction. Using this, Ezung et al. \cite{egkp} have obtained a part of the excited state spectrum for the JK model. Proceeding along similar lines, we begin by writing wavefunctions of the excited states in the form 
\be\label{10}
    \psi = \psi _0 \phi,
\ee
where $\psi_0$ is the ground state wavefunction as given by \eqref{4} and $\phi$ is required to be symmetric so that $\psi$ behaves like $\psi_0$ under the exchange of particles. Plugging this $\psi$ into our Hamiltonian \eqref{5}, we get
\be\label{11}
    H_1 \phi = \left(\epsilon - \epsilon_0\right) \phi,
\ee
where $H_1$ is given by
\be\label{12}
    H_1 = -\frac{1}{2} \sum_{j=1}^{N} \frac{\partial^2}{\partial x_j^2} + \beta \frac{\pi}{L} \sum_{\substack{k =1}}^ {r} \sum_{\substack{j =1}}^ {N} \bigg(\cot \left[\frac{\pi}{L}\left(x_j-x_{j-k}\right)\right] - \cot \left[\frac{\pi}{L}\left(x_j-x_{j+k}\right)\right] \bigg),
\ee
while $\epsilon -\epsilon_0 = (E-E_0)(L/2\pi)^2$. In order to obtain the excited state solutions we see that it is suitable to use the variables
\begin{alignat}{1}\label{13}
    &z_j = e^{2\pi \iota x_j / L}, \nonumber \\  &\text{giving} ~ \cot \left[\frac{\pi}{L}\left(x_j-x_{j+k}\right)\right] = \iota \frac{z_j-z_{j+k}}{z_j-z_{j+k}},
\end{alignat} 
and hence, $H_1$ takes the form
\be\label{14}
    H_1 = \sum_{j=1}^{N} D_j^2 + \beta \sum_{\substack{k =1}}^{r} \sum_{\substack{j =1}}^ {N} \frac{z_j-z_{j+k}}{z_j-z_{j+k}} \left(D_j - D_{j+k}\right),
\ee
where $D_j = z_j \frac{\partial}{\partial z_j} $. \\
Some remarks concerning certain interesting features of \eqref{11} and \eqref{14} are in order:
\begin{enumerate}
    \item $H_1$ commutes with the momentum operator $P = \frac{2 \pi}{L}\sum_{i=1}^{N} z_i \frac{\partial}{\partial z_i}$. Thus if $\phi$ is eigenfunction of $H_1$, it is also eigenfunction of $P$, i.e., 
    \be\label{15}
    P \phi = \kappa \phi.
    \ee

    \item If $\phi$ is an eigenfunction of $H_1$ and $P$, then $\phi'$ given by 
    \be\label{16}
    \phi' = G^{q} \phi,~~ G = \prod_{i=1}^{N} z_i,
    \ee 
    is also an eigenstate of $H_1$ and $P$ with eigenvalues $\epsilon - \epsilon_0 + 2Nq(L/2\pi)\kappa +(N q)^2$ and $\kappa + N q $ respectively. Note that here $q$ is any integer (both positive and negative). Observe that the multiplication by $G$ implements a Galilei boost.

    \item The Hamiltonian $H_1$ and hence the eigenvalue \eqref{11} are invariant under $z_j \rightarrow z_{j}^{-1}$. Since $z_j = e^{2i\pi x_j/L}$, hence $z_{j}^{-1} = e^{-2i \pi x_j/L}$ thereby indicating the presence of the left as well as right moving modes with eigenvalues $\kappa$ and $-\kappa$ respectively. Hence it follows that if there is a solution with momentum $\kappa$ then there must exist another solution with the same energy but momentum -$\kappa$. Thus all the excited states with nonzero momentum are (at least) doubly degenerate.
\end{enumerate}

Let us now discuss the excited state solutions by using \eqref{11} and \eqref{14}. It is easily shown that $H_1$ as given by \eqref{14} admits four excited state solutions with the corresponding eigenvalues and eigenfunctions:
\be\label{17}
\begin{split}
\phi &= e_1, \quad \epsilon-\epsilon_0 = 1+2r\beta  \\
\phi &= e_{N-1}, \quad \epsilon-\epsilon_0 = (N-1)+2r\beta  \\
\phi &= e_N, \quad \epsilon-\epsilon_0 = N  \\
\phi &= e_1 e_{N-1}- \frac{N}{1+2r\beta} e_N, \quad \epsilon-\epsilon_0 = N+2(1+2r\beta)
\end{split}
\ee

Here $e_j \, (j=1,2, \dots , N)$ denotes elementary symmetric functions in terms of $z_j$. For instance, $e_2 = z_1 z_2 + z_2 z_3+ \dots + z_{N-1} z_N$ and has $N(N-1)/2$ number of terms. It is interesting to note that the energy of the third state in the above list is independent of the range of interactions $r$. In the appropriate limit (i.e. $r =1$) we recover the four known excited state energies for the JK model. 

As  mentioned above, each of these solution is doubly degenerate. For example, the two solutions $e_1$ and $e_{N-1}/e_N$ are degenerate. By taking the linear combination of these two complex solutions one can show that the two degenerate real solutions are
\begin{equation}
\phi = \sum_{i=1}^{N} \cos u_i,\quad \phi = \sum_{i=1}^{N} \sin u_i
\end{equation}
where $u_i = 2\pi x_i/L$. Similarly, all other degenerate excited state solutions may be rewritten as two independent real solutions. Despite the apparentness all the excited are not doubly degenerate. In particular, consider
\begin{equation}
\phi = \frac{e_1 e_{N-1}}{e_N}  - \frac{N}{(1+2r\beta)}.
\end{equation}
Rewritten in terms of trigonometric functions, it becomes
\begin{equation}
\sum_{i<j}^{N} \cos(u_i -u_j) + \frac{N \beta}{(1+2r\beta)}.
\end{equation}
This is an exact solution with $\epsilon - \epsilon_0 = 2+4 r \beta$ but with momentum eigenvalue $\kappa = 0$. And it is a non-degenerate solution as it remains invariant under $z_i \rightarrow z_{i}^{-1}$. 

We would like to point out that apart from the aforementioned four states, the degeneracy structure and other excited eigenstates, if they exist, still remain unknown.

\section{Conclusions}
We have presented here a new class of models that describe particles confined to a circle and subjected to inverse-square, two-body and three-body interactions among a finite number of neighbors. This interpolates between the Sutherland model on one hand (when the three-body terms vanish) and the JK model on the other. The exact ground state energy and a few excited states have been determined.

The variable range of interaction, we believe, is closer to physical situations where screening is present and therefore, experimental realizations of such one dimensional models would be interesting to see. And with impetus from the advances in field of ultracold atoms \cite{KWW} that may not be too far away.

This family of truncated range models can be extended in a wide variety of ways. For example one could extend it in the case of multiple species, in higher dimensions and to a variety of root systems. Further, one can inquire if like JK model, these models also have off-diagonal long range order (ODLRO). It may be recalled that the JK model is the only known one-dimensional model that exhibits ODLRO, implying Bose-Einstein condensation at zero temperature \cite{jk1}. Besides, one can try to map this problem to some circular random matrix model. For the Sutherland and JK models, there is a transformation which leads one to the exact eigenfunctions for classically chaotic systems \cite{jgk}, a similar mapping makes a connection with a five-dimensional billiard for the model presented here. We hope to return to some of these connections in the near future.

\section{Acknowledgments}
This work was carried out while one of the authors (TRT) was visiting Bhabha Atomic Research Centre (BARC), Mumbai as a part of a programme jointly organized by the Indian Academy of Sciences, National Academy of Sciences of India, and the Indian National Science Academy. TRT thanks the Academies and BARC for the support extended during the stay. AK is grateful to Indian National Science Academy (INSA) for the award of INSA Senior Scientist position at Savitribai Phule Pune University.

\end{document}